\documentclass[twocolumn, times, tighten]{aastex63}
\usepackage{graphicx,multirow,hyperref,url,color,xspace}

\newcommand{\msun}{{\rm M}_{\sun}}

\newcommand{\rg}{{R_\mathrm{g}}}

\received{2020 February 27}
\accepted{2020 April 23}

\begin{document}

\title{The persistent radio jet coupled to hard X-rays in the soft state of Cyg X-1}
\shorttitle{The soft state radio jet of Cyg X-1}

\author{Andrzej A. Zdziarski}
\affil{Nicolaus Copernicus Astronomical Center, Polish Academy of Sciences, Bartycka 18, PL-00-716 Warszawa, Poland; \href{mailto:aaz@camk.edu.pl}{aaz@camk.edu.pl}}

\author{J. N. S. Shapopi}
\affil{Department of Physics, University of Namibia, Windhoek, Namibia; \href{mailto:jimzashapopi@yahoo.com}{jimzashapopi@yahoo.com}}

\author{Guy G. Pooley}
\affil{Cavendish Laboratory, J. J. Thomson Avenue, Cambridge CB3 0HE, UK }

\shortauthors{Zdziarski et al.}

\begin{abstract}
We study long-term radio/X-ray correlations in Cyg X-1. We find the persistent existence of a compact radio jet in its soft state. This represents a new phenomenon in black-hole binaries, in addition to compact jets in the hard state and episodic ejections of ballistic blobs in the intermediate state. While the radio emission in the hard state is strongly correlated with both the soft and hard X-rays, the radio flux in the soft state is not directly correlated with the flux of the dominant disk blackbody in soft X-rays, but instead it is lagged by about a hundred days. We interpret the lag as occurring in the process of advection of the magnetic flux from the donor through the accretion disk. On the other hand, the soft-state radio flux is very tightly correlated with the hard X-ray, 15--50\,keV, flux without a measurable lag and at the same rms. This implies that the X-ray emitting disk corona and the soft-state jet are powered by the same process, probably magnetically.
\end{abstract}

\section{Introduction}
\label{intro}

Our current knowledge of the radio emission of accreting black-hole (BH) binaries includes power-law correlations between the soft/medium X-rays and the radio flux emitted by steady compact jets in the hard spectral state, episodic ejection of radio-emitting blobs in the intermediate state, and jet quenching in the soft state (e.g., \citealt{FBG04,Fender09, Miller-Jones12, Corbel13, Kalemci13}). In some works \citep{ZSPL11,Islam18,Koljonen19}, the correlations with the bolometric luminosity (including hard X-rays) were studied, but no significant differences with respect to the previous studies were found in the hard state. The radio emission of the compact jets is partially synchrotron self-absorbed (with the energy spectral index of $\alpha\sim0$, e.g., \citealt{Fender00}), while that of the episodic jets is usually optically thin (e.g., \citealt{Rodriguez95}). Both types of jets are probably mildly relativistic. The two main proposed mechanisms for the formation of either type utilize large scale magnetic field co-rotating with either a Kerr BH \citep{BZ77} or an accretion disk \citep{BP82}. 

In the case of the process of \citet{BZ77} and saturated large-scale magnetic field ('magnetically arrested' or 'choked', \citealt{BK74, Narayan03, McKinney12}), the jet power was shown to be capable of reaching $\sim\!\!\!a_*^2\dot{M}c^2$, where $\dot{M}$ and $a_*$ are the accretion rate and the BH spin parameter, respectively \citep{Tchekhovskoy11}. An approximate dependence on $a_*^2$ was claimed for episodic jets \citep{Narayan12}, but none was found for steady compact jets of binaries \citep{Fender10}.

The jet power may obviously be lower than $a_*^2\dot{M}c^2$, but it is still expected to be proportional to the square of the magnetic flux in the BH vicinity. An important issue is the origin of the required large-scale magnetic field. In accreting binaries, it could be advected from the donor star \citep{BK74}. This process is efficient in geometrically-thick, hot, disks, but very inefficient in standard geometrically thin disks because of the magnetic diffusion \citep{Lubow94}. This appears to be consistent with the jet quenching in the soft state of BH binaries. However, it cannot explain the presence of jets in the hard state, given that their hot disks are present only close to the BH and are surrounded by large thin disks (e.g., \citealt{DHL01}). Those jets may be possibly generated by a local process, e.g., \citet{Liska18}.

Here, we study long-term radio/X-ray correlations in the bright BH binary Cyg X-1, with the goal of testing the above paradigms. (These correlations are also discussed in \citealt{Shapopi20}.) This system accretes via stellar wind from an OB supergiant, its X-ray emission originates from accretion, while the radio emission is from the jet, resolved by VLBA and VLA observations \citep{Stirling01}. In the hard state, the spectrum is dominated by a hard power law with a high energy cut-off, well modelled by thermal Comptonization \citep{G97}. In the soft state, the blackbody disk emission dominates the spectrum, and it is followed by a high energy tail, which is well fitted by Comptonization by electrons with a hybrid distribution with a significant non-thermal tail \citep{Gierlinski99}. 

\section{The Data and Analysis Method}
\label{data}

\setlength{\tabcolsep}{3.5pt}
\begin{table}[t]
\centering            
\caption{The adopted inclusive MJD intervals of the occurrences of the hard/intermediate state.}
\begin{tabular}{lccccccc}
\hline
Start & End & Start & End & Start & End & Start & End\\
\hline
50085 &50222 &52853 &53003 &55895 &55940 &57105 &57265\\
50308 &51845 &53025 &53265 &56035 &56087 &57332 &57970\\
51858 &52167 &53292 &53368 &56722 &56748\\
52205 &52237 &53385 &55387 &56760 &56845\\
52545 &52801 &55674 &55790 &57012 &57045\\
\hline
\end{tabular}
\label{dates}
\end{table} 

\begin{figure}
  \centering
\includegraphics[width=7cm]{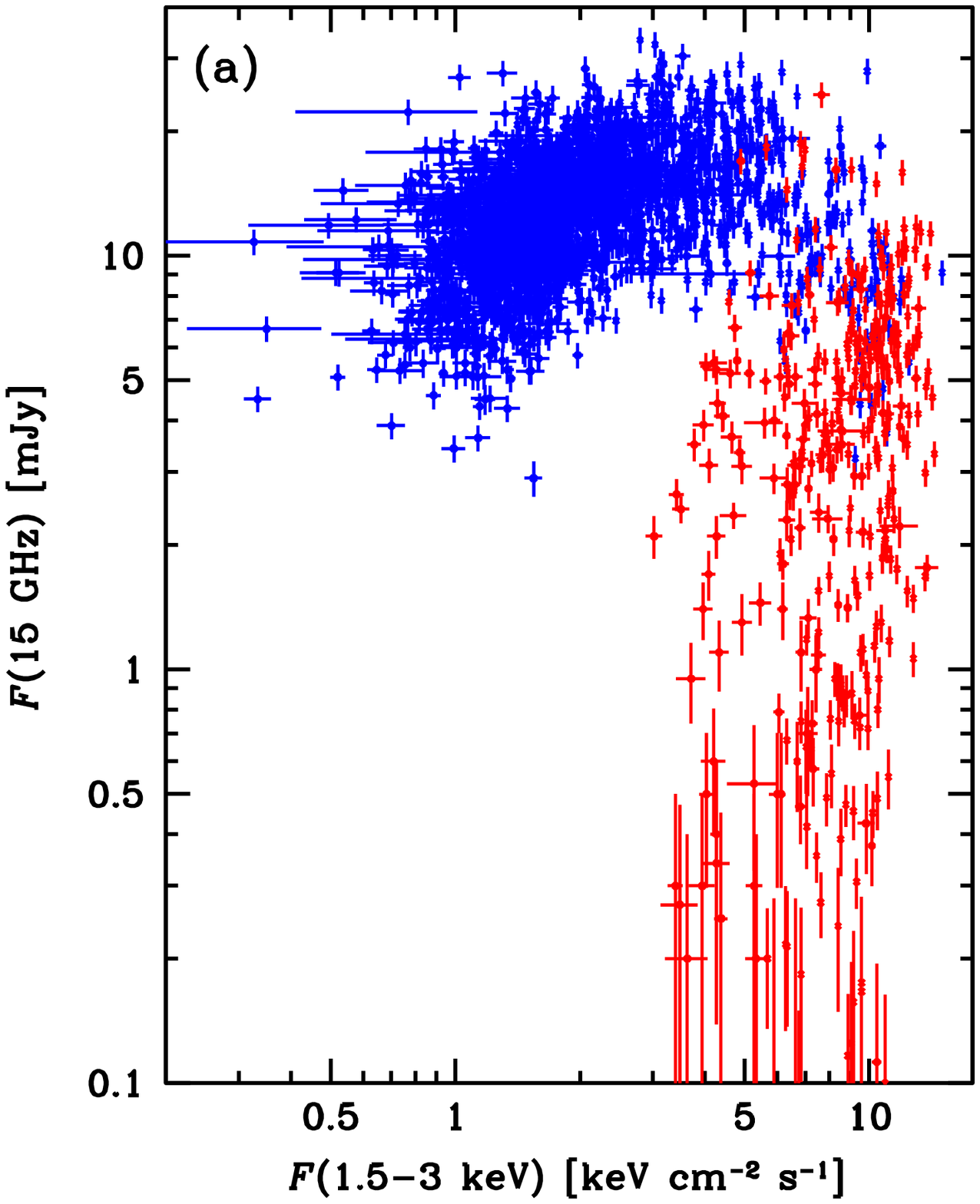}  \includegraphics[width=7.cm]{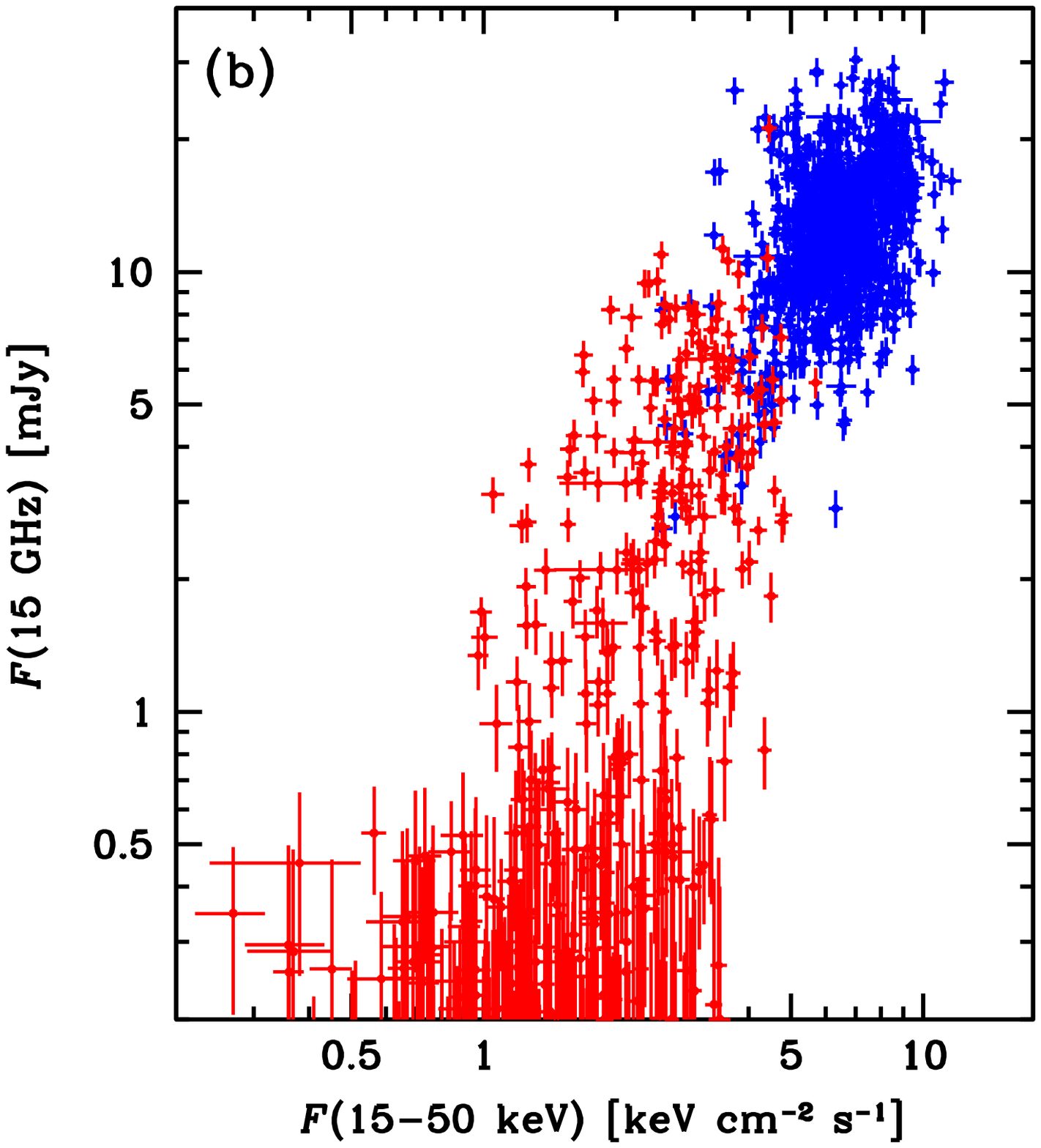}
  \caption{The correlations between the daily average 15\,GHz flux and (a) the 1.5--3\,keV and (b) 15--50\,keV fluxes (on the same day). The blue and red points correspond to the hard/intermediate and soft states, respectively, following Table \ref{dates}. Only fluxes with the fractional error $<0.5$ are plotted.
}
\label{zero_lag}
\end{figure}

In soft/medium X-rays, we use the light curve for MJD 50087--55870 from the All Sky Monitor (ASM; \citealt{ASM}) on board of {\it Rossi X-ray Timing Explorer}, which covers three bands, 1.5--3, 3--5 and 5--12\,keV. In hard X-rays, we use the 15--50\,keV light curve for MJD 53416--58000 from the Burst Alert Telescope (BAT; \citealt{BAT}) on board of {\it Neil Gehrels Swift}. We correlate the radio and X-ray light curves either separately for different spectral states or for all of the data. In addition to the hard and soft states, Cyg X-1 has a distinct intermediate state, and we separate the states following \citet{Grinberg13} and \citet{ZMC17}. For that, we also use the 2--4, 4--10 and 10--20\,keV data from the Monitor of All-sky X-ray Image (MAXI; \citealt{Matsuoka09}). Cyg X-1 spends most of the time in either the hard or soft states and much less in the intermediate state (e.g., \citealt{Grinberg13}); thus the number of observations in the latter is small. Since the radio/X-ray correlation in Cyg X-1 has similar properties in both hard and intermediate states \citep{ZSPL11}, we treat them jointly. (We also have tested joining the soft and intermediate states, and have obtained no qualitative difference in our results.) We convert the X-ray count rates into physical fluxes using an absorbed power law model for the Crab, and calculate the hard X-ray photon spectral index, $\Gamma$, based on either the MAXI 2--10 keV range or the ASM 3--12 keV following the method of \citet{ZPS11}. We set the boundary between the soft and the intermediate states at $\Gamma\approx2.4$--2.5 \citep{Grinberg13, ZMC17}. This gives us the hard/intermediate days listed in Table \ref{dates}. Below, the term `hard state' denotes the sum of the hard and intermediate states.

\begin{figure}
  \centering
  \includegraphics[width=7.cm]{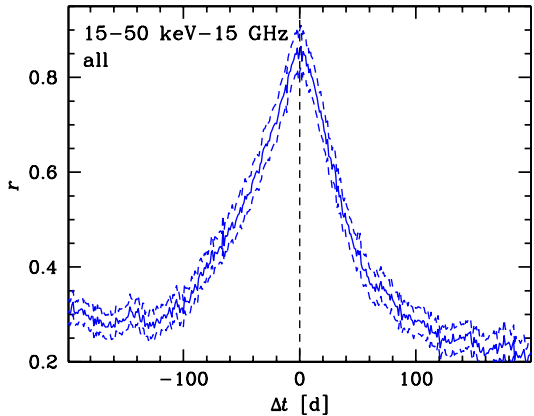}
  \caption{The 15\,GHz vs.\ 15--50\,keV cross-correlation for the entire light curves, showing the correlation coefficient of close to unity at zero lag. Hereafter, the solid lines show the calculated values of $r$ while the dashed curves outline the statistical uncertainty region.
}\label{r_hardx}
\end{figure}

\begin{figure*}
\centerline{\includegraphics[width=6.cm]{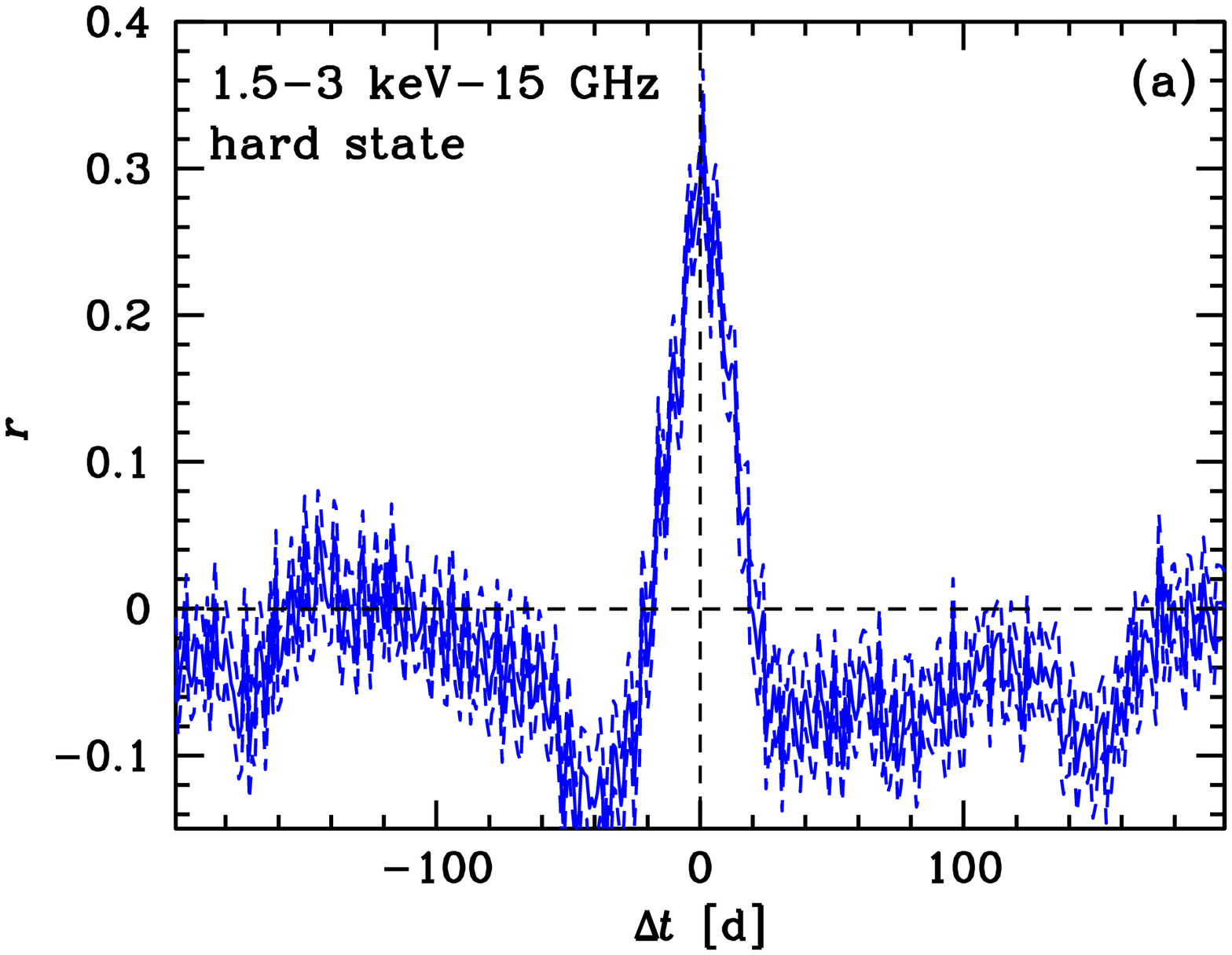}
\includegraphics[width=6.cm]{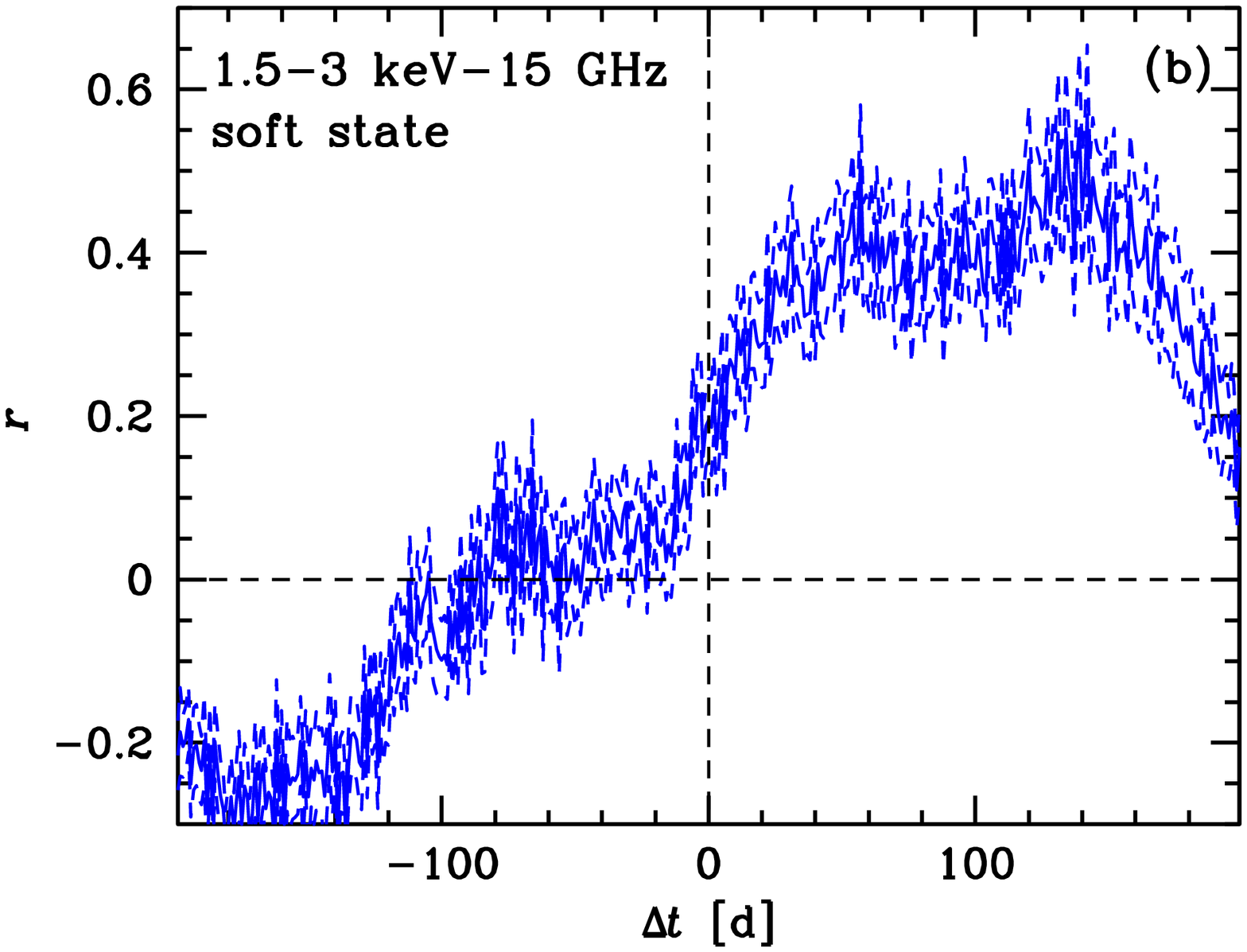}
\includegraphics[width=6.cm]{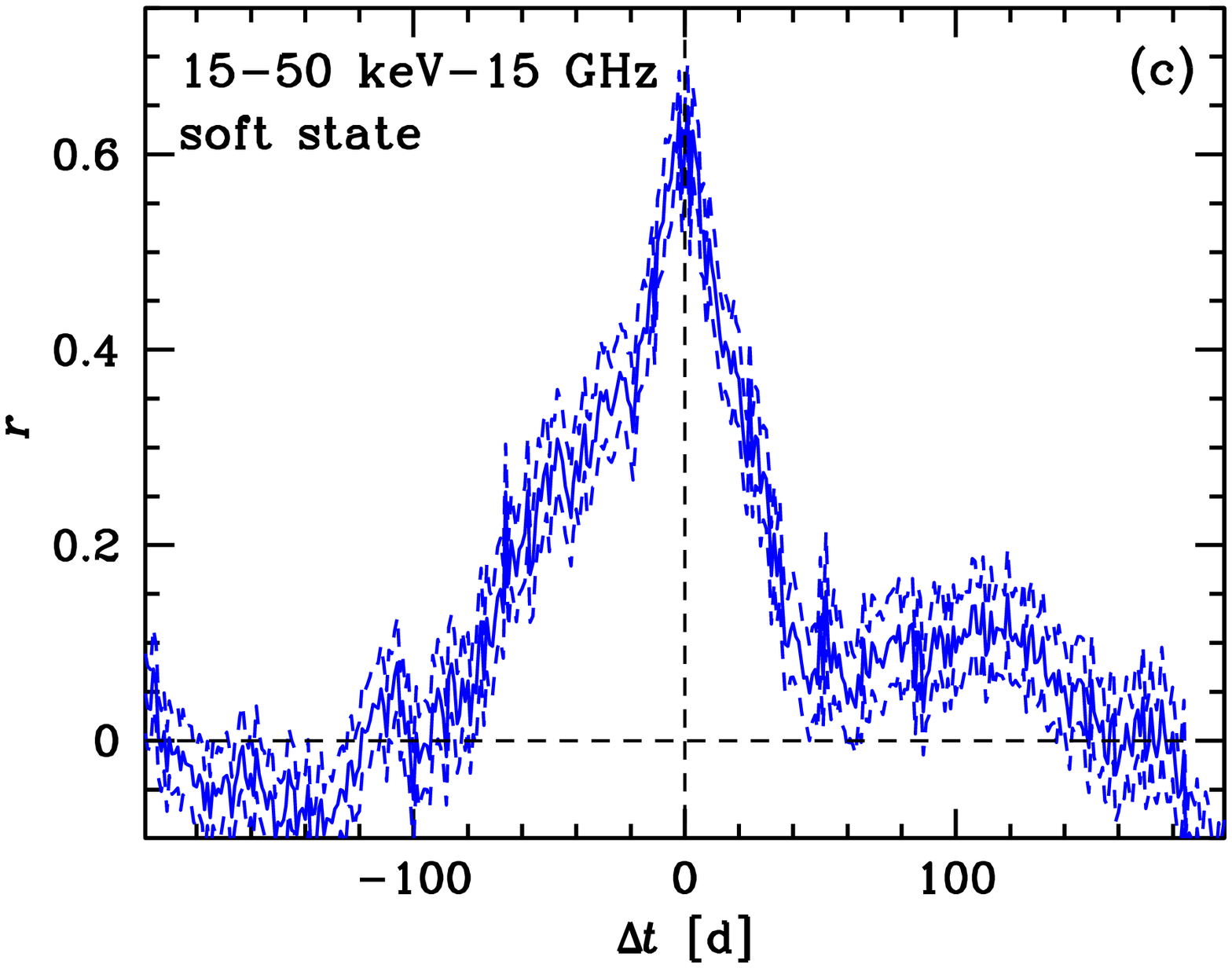}}
  \caption{The 15\,GHz vs X-ray cross-correlation for (a) 1.5--3\,keV in the hard state, and (b) 1.5--3\,keV, (c) 15--50\,keV in the soft state. We see a narrow zero-lag correlation in the hard state in the soft X-rays, similar to that for all the data in the hard X-rays (Fig.\ \ref{r_hardx}). In the soft state, we see a strong evolution of the radio emission from lagging the 1.5--3\,keV X-rays by $\sim$50--150\,d to a strong peak at zero lag in 15--50\,keV X-rays.
}\label{hard_soft}
\end{figure*}

We use the radio monitoring light curves at 15\,GHz for MJD 50226--53902 from the Ryle telescope \citep{PF97}, and for MJD 54573--57748 from the Arcminute Microkelvin Imager (AMI; \citealt{Zwart08,Hickish18}), as published in \citet{ZMC17}. The individual pointings have statistical errors $\lesssim$0.2\,mJy, and are subject to variations in the flux calibration of $\lesssim7\%$ from one day to another (Dave Green, private communication). We assign both errors added in quadrature to the fluxes of individual pointings. For daily averages, we add in quadrature the statistical error of the average to the 7\% daily variation.

The cross-correlation coefficients are calculated using the method of \citet{EK88} with some modifications described in \citet{Z18}, where the Pearson's correlation coefficient, $r$, for two discrete and unevenly spaced light curves shifted in time by $\Delta{t}$ is calculated using the logarithms of the fluxes. In our convention, $\Delta{t}>0$ corresponds to radio lagging X-rays. We follow a few variants of this method, see a discussion in \citet{Z18}. In particular, we use either the individual detections or daily averages, but find the results using different variants are relatively similar, and show here those for the daily averages. We take into account the fluxes with the relative error $<1$. We normalize the cross-correlation coefficient to the square roots of the variances of the data sets. We do not subtract the average measurement variance from the total, which is a small effect for our data sets. Including it would slightly increase the reported values of $r$. Given the time resolution and relatively low sensitivity of the observations, we cannot meaningfully find lags on a time scale less than a day; we thus use the term 'zero lag' for the bin at $0\pm0.5$\,d. Here, we correlate X-ray photons from within a given state interval with the radio photons from the entire light curve, which allows us to include those emitted after the end of that interval. 

\section{Radio/X-ray correlations}
\label{corr}

Figures \ref{zero_lag}a and b show the fluxes in the radio band vs.\ those of the soft, 1.5--3\,keV and hard, 15--50\,keV X-rays, respectively, for the daily averages of the ASM, BAT and 15\,GHz data. Figure \ref{zero_lag}a shows a positive flux-flux radio/1.5--3\,keV X-ray correlation in the hard state, and almost no correlation in the soft state. Figure \ref{zero_lag}b shows that the radio and 15--50\,keV X-rays form a remarkable single track across both the hard and soft states. This single track is reflected by the radio/15--50\,keV cross-correlation including all the states of Cyg X-1 peaking at zero lag with a very high correlation coefficient, $r\approx0.86\pm0.04$, as shown in Figure \ref{r_hardx}. The cross-correlation has the half-width of $\approx$60\,d.

Figures \ref{hard_soft}a, b show the correlations of the radio with the softest ASM band, 1.5--3\,keV, in the hard and soft states, respectively. In the hard state, we find the cross-correlation peaking at zero lag, similarly to the case shown in Figure \ref{r_hardx}. However, we find a strongly delayed radio emission in the soft state, Figure \ref{hard_soft}b. While the correlation at zero lag is weak, $r\approx0.2$ (in agreement with Figure \ref{zero_lag}a), it increases with the increasing positive lag up to the broad maximum of $r\approx0.5$--0.6 at $\Delta{t}\sim50$--150\,d. This delayed maximum in the soft state weakens to $r\approx0.3$--0.4 in the 3--5\,keV range, while the 5--12\,keV band shows the dominant peak at zero lag (not shown here). If we combine the intermediate and soft states (for the boundary with the hard state at $\Gamma\approx2$), the results are similar but the maximum values of $r$ are lower. In the 15--50\,keV band, Figure \ref{hard_soft}c, we see a strong single peak at zero lag. This clearly shows the different nature of the soft and hard X-rays in the soft state. While the peak values of the cross-correlation are similar for the 1.5--3 and 15--50\,keV bands, $r\approx0.6$, the former occurs at $\Delta{t}\approx140$\,d and the latter is at $\Delta{t}\approx0$\,d. 

Figure \ref{hard_soft} also shows low-amplitude periodic modulation of the cross-correlation at the orbital period of 5.6\,d, which, on the scale of the figures, appears as a widening of the $r$ profile. It is caused by the modulated absorption of both radio \citep{Zdziarski12} and soft X-rays \citep{Wen99} in the wind of the donor. The source also shows superorbital quasi-periodic modulation seen in the hard state only, with the quasi-period of $\sim$150\,d until around MJD 53500, and $\sim$300\,d after that date, e.g., \citet{Poutanen08}, \citet{ZPS11}. This could, in principle, introduce repetitive features in the $\Delta{t}$-$r$ space with the periods equal to those of the modulation (as in the case of the orbital modulation). However, no such features are seen in the hard state, Figure \ref{hard_soft}a, and Figure \ref{hard_soft}b (the soft state) shows the $\sim$100\,d lag only for $\Delta{t}>0$. Thus, we rule out the possibility of the radio lag in the soft state being an artefact of the superorbital modulation.

\section{Discussion}

Our statistically strongest result is the very tight correlation between the 15\,GHz and 15--50\,keV fluxes across all of the states of Cyg X-1, with the correlation coefficient of $\approx$0.86$\pm0.04$ at zero lag. This shows the jet emission is strongly linked to the hot Comptonizing medium, either a hot disk or a hot corona. The correlation in the hard state agrees with our standard knowledge of that state, featuring compact steady jets. However, the radio emission in the soft state of Cyg X-1 represents {\it a new phenomenon} in BH binaries. It is clearly not from episodic ballistic jets, occurring in the intermediate state (e.g., \citealt{FBG04, Miller-Jones12}); instead it is from a small compact jet \citep{Rushton10,Rushton12}, which persists through the soft state. Also, the jet emission of Cyg X-1 in the soft state remained partially synchrotron self-absorbed (with $\alpha\approx0$) during the few available 4--13\,GHz measurements at fluxes $\gtrsim1$\,mJy \citep{Rushton10}. 

While the soft-state radio emission is highly variable, it remains tightly connected to the similarly variable soft-state hard X-rays. We have calculated the fractional rms variability of the used 15--50\,keV and 15\,GHz soft-state light curves, and found them to be $0.52\pm0.01$ and $0.55\pm0.01$, respectively, i.e., practically identical. Thus, the variability of the hot corona appears to be transferred to the jet without any loss. On the other hand, the rms of the 1.5--3\,keV light curve is significantly lower, $0.28\pm0.01$. The hard X-rays are clearly from Comptonization of the disk blackbody in a corona \citep{Gierlinski99}, probably powered magnetically. The spectra from this process cover a range significantly broader than 15--50\,keV, and the rms increases with the energy in the soft and intermediate states. Thus, the rms of the total power supplied to the corona could be slightly different from that measured in the BAT range. The corona is likely to represent the base of the jet. However, the hard X-ray emission, which has a steep spectrum, $\Gamma\gtrsim2.5$, cannot be an optically-thin extension of the radio synchrotron spectrum, as that would imply, when extrapolated to putative synchrotron break energy, unrealistically high powers.

An important difference between the compact jet radio emission in the hard and soft states of Cyg X-1 is that while the former is correlated with the bolometric luminosity, $L$, and then $\dot{M}$ \citep{ZSPL11}, the soft state emission is not correlated at all with the instantaneous $\dot{M}$, given that the hard X-rays in that state represent a minor fraction of $L$. On the other hand, the correlation with $\dot{M}$ can be recovered if the radio lag we have found with respect to the soft X-rays is taken into account. The soft X-rays are from the blackbody emission of the optically-thick disk extending to the innermost stable circular orbit, and they do trace the bolometric $L$. Thus, the jet power appears to be still related to $\dot{M}$, but at an earlier epoch. 

We interpret the lag as the time between a fluctuation in the thin accretion disk arriving near the black hole and the subsequent response of the jet. The jet production in a thin disk requires the presence of poloidal magnetic field \citep{Liska19}, and this field can be advected from the donor star. Such advection may be delayed by some process. A similar ($\sim$50\,d) radio lag with respect to the soft X-rays in the soft state was found in Cyg X-3 \citep{Z18}, and the mechanism for field advection in Cyg X-1 can be similar to that proposed by \citet{CZ20}, in which it is due to magnetically-driven disk outflows facilitating the field advection. The outflows occur only after certain threshold accretion rate is reached. The characteristic time scale for the delay in this process is the viscous time scale, $t_\mathrm{visc}$, at the disk outer edge, $R_\mathrm{out}$. \citet{SL76} estimated $R_\mathrm{out}\sim\!10^3\rg$ in Cyg X-1 ($\rg\equiv{GM}_\mathrm{BH}/c^2$, where $M_\mathrm{BH}$ is the BH mass), at which radius $t_\mathrm{visc}$ is of the order of days. However, the disk formation in systems like Cyg X-1, where the donor almost entirely fills the Roche lobe and the wind is strongly focused, is likely to be significantly different from nearly-spherical wind accretors, and the actual outer radius can be larger. Thus, we estimate the outer radius that would correspond to a given viscous time scale,
\begin{equation}
R_\mathrm{out}= (\alpha t_\mathrm{visc})^{2/3}(G M_\mathrm{BH})^{1/3} (H/R)^{4/3},
\label{radius}
\end{equation}
where $\alpha\sim0.1$ is the viscosity parameter and $H/R\sim0.1$ is the fractional disk scale height. At $t_\mathrm{visc}=100$\,d, $M_{\rm BH}=20\msun$ \citep{Ziolkowski14} and those values of $\alpha$ and $H/R$, this yields $R_\mathrm{out}\approx6\times10^{11}{\rm cm}\approx2\times10^5\rg$ (with a large uncertainty), which is about a half of the tidal radius. We consider it a plausible value. While the time scale of state transitions in Cyg X-1 is much shorter than this $t_\mathrm{visc}(R_\mathrm{out})$, it is similar to those in low-mass X-ray binaries (LMXBs), where $R_\mathrm{out}$ is certainly near the tidal radius. This shows that state transitions are due to $\dot{M}$ changes occurring much closer to the compact object than the outer disk radius.

We find significant similarities between the long-term radio/X-ray correlations in Cyg X-1 and Cyg X-3. In particular, the radio emission of the latter is tightly correlated at zero lag with the hard X-rays in all states but not with the soft X-rays. The radio emission is lagged by $\sim$50\,d with respect to soft X-rays in the soft state of Cyg X-3 \citep{Z18}, while this lag is $\sim$150\,d in Cyg X-1, which difference can be attributed to the larger binary separation of Cyg X-1. One difference is that the hard X-rays in the hard state are anti-correlated with the radio emission in Cyg X-3 (e.g., \citealt{Z18}), while they are positively correlated in Cyg X-1. This can be explained by the pivoting variability with the pivot energy being between soft and hard X-rays in Cyg X-3. While the soft-state radio emission of Cyg X-3 can be much more powerful than that in its hard state, and then coming from a spatially resolved jet, its size is still $\lesssim$ a few times $10^{15}$\,cm (e.g., \citealt{Egron17}), which is relatively similar to the size of the compact jet in Cyg X-1. Also, that emission of Cyg X-1 is relatively steady and does not represent episodic ejections.

It is then of interest whether the found behavior of the soft state of Cyg X-1 occurs also in BH LMXBs. We should take into account that the former shows significant high-energy tails (see fig.\ 4a in \citealt{ZG04}), while there are usually relatively weak high-energy tails beyond the disk blackbody in the soft state of LMXBs (see figs.\ 8 and 9a in \citealt{ZG04}). Still, the soft state of both Cyg X-1 and BH LMXBs share the same basic features. One is the presence of a strong and stable (on time scales $\lesssim$ hours) disk component, cf.\ \citet{CGR01} vs.\ \citet{GD04}. Another is the similarly high levels of variability of high-energy spectral tails, e.g., \citet{Gierlinski10} vs.\ \citet{GZ05}. Thus, the lack of detectable radio emission in the soft state of BH LMXBs could be due to the relative weakness of their high-energy tails, and still in agreement with the radio correlation with hard X-rays in Cyg X-1. In order to test it, we should then look at transitional states of LMXBs, which do have significant high-energy tails. The turning-on of jets during the soft-to-hard transitions during the outburst decays has been studied by, e.g., \citet{Kalemci13}, but without looking at the relationship to hard X-rays. \citet{Kubota_Done16} studied a transitional very high state of the BH LMXB GX 339--4 accompanied by weak radio detection. They stated that the radio flux was weaker than that expected based on the flux of the Comptonization component, but this represents just a single point in the putative correlation. Then, \citet{Fender09} detected weak radio emission in a number of BH LMXBs in the soft state, but they found it to be consistent with the origin in jet-ISM interactions far from the BH, with the core radio emission switched off. Clearly, more studies of the radio emission from the soft and transitional states of BH LMXBs are desirable.

Finally, we note a significant similarity of the soft-state radio emission in Cyg X-1 with the radio emission of blazars with most powerful jets, which are usually found in disk-dominated, soft, states of radio-loud active galactic nuclei, see, e.g., the sample of \citet{Zamaninasab14}. The required large-scale magnetic field is then thought to be advected from coherent patches of the ISM (e.g., \citealt{Cao19}), analogously to the field advection from the donor that we postulate to take place in Cyg X-1.

\section{Conclusions}

We have established the persistent existence of a compact jet in the soft state of Cyg X-1. This represents a new jet phenomenon in BH binaries, in addition to compact jets in the hard state, and episodic ejections of ballistic blobs in the intermediate state. The radio flux in the soft state of Cyg X-1 is tightly correlated with the hard X-ray flux, with both being strong variable at similar values of the rms of $\approx$50\%. 

While the radio flux in the soft state is not correlated (at zero lag) with the soft X-ray flux of the dominant disk blackbody, we have discovered that the former lags the latter with $\Delta{t}\sim50$--150\,d. We interpret the lag as occurring in the process of advection of the magnetic flux from the donor through the accretion disk. The tight soft-state radio/hard X-ray correlation then indicates that both the X-ray emitting disk corona and the jet are powered by the same process, most likely magnetic.

This type of radio emission in Cyg X-1 shows similarities to emission of Cyg X-3, as well as to the jets in soft accretion states of blazars. On the other hand, the relationship to BH LMXBs remains unclear, and we stress the importance of studying the radio/hard X-ray correlation in their soft and transitional states.

\section*{Acknowledgments}
We thank Barbara De Marco for a lot of valuable comments, and Dave Green and Emrah Kalemci for valuable discussions. We also acknowledge valuable comments of the referee, and support from the Polish National Science Centre under the grant 2015/18/A/ST9/00746.


\begin{thebibliography}{}
\expandafter\ifx\csname natexlab\endcsname\relax\def\natexlab#1{#1}\fi
\providecommand{\url}[1]{\href{#1}{#1}}
\providecommand{\dodoi}[1]{doi:~\href{http://doi.org/#1}{\nolinkurl{#1}}}
\providecommand{\doeprint}[1]{\href{http://ascl.net/#1}{\nolinkurl{http://ascl.net/#1}}}
\providecommand{\doarXiv}[1]{\href{https://arxiv.org/abs/#1}{\nolinkurl{https://arxiv.org/abs/#1}}}

\bibitem[{{Barthelmy} {et~al.}(2005){Barthelmy}, {Barbier}, {Cummings},
  {Fenimore}, {Gehrels}, {Hullinger}, {Krimm}, {Markwardt}, {Palmer},
  {Parsons}, {Sato}, {Suzuki}, {Takahashi}, {Tashiro}, \& {Tueller}}]{BAT}
{Barthelmy}, S.~D., {Barbier}, L.~M., {Cummings}, J.~R., {et~al.} 2005, \ssr,
  120, 143, \dodoi{10.1007/s11214-005-5096-3}

\bibitem[{{Bisnovatyi-Kogan} \& {Ruzmaikin}(1974)}]{BK74}
{Bisnovatyi-Kogan}, G.~S., \& {Ruzmaikin}, A.~A. 1974, \apss, 28, 45,
  \dodoi{10.1007/BF00642237}

\bibitem[{{Blandford} \& {Payne}(1982)}]{BP82}
{Blandford}, R.~D., \& {Payne}, D.~G. 1982, \mnras, 199, 883,
  \dodoi{10.1093/mnras/199.4.883}

\bibitem[{{Blandford} \& {Znajek}(1977)}]{BZ77}
{Blandford}, R.~D., \& {Znajek}, R.~L. 1977, \mnras, 179, 433,
  \dodoi{10.1093/mnras/179.3.433}

\bibitem[{{Cao} \& {Lai}(2019)}]{Cao19}
{Cao}, X., \& {Lai}, D. 2019, \mnras, 485, 1916, \dodoi{10.1093/mnras/stz580}

\bibitem[{{Cao} \& {Zdziarski}(2020)}]{CZ20}
{Cao}, X., \& {Zdziarski}, A.~A. 2020, \mnras, 492, 223,
  \dodoi{10.1093/mnras/stz3447}

\bibitem[{{Churazov} {et~al.}(2001){Churazov}, {Gilfanov}, \&
  {Revnivtsev}}]{CGR01}
{Churazov}, E., {Gilfanov}, M., \& {Revnivtsev}, M. 2001, \mnras, 321, 759,
  \dodoi{10.1046/j.1365-8711.2001.04056.x}

\bibitem[{{Corbel} {et~al.}(2013){Corbel}, {Coriat}, {Brocksopp}, {Tzioumis},
  {Fender}, {Tomsick}, {Buxton}, \& {Bailyn}}]{Corbel13}
{Corbel}, S., {Coriat}, M., {Brocksopp}, C., {et~al.} 2013, \mnras, 428, 2500,
  \dodoi{10.1093/mnras/sts215}

\bibitem[{{Dubus} {et~al.}(2001){Dubus}, {Hameury}, \& {Lasota}}]{DHL01}
{Dubus}, G., {Hameury}, J.-M., \& {Lasota}, J.-P. 2001, \aap, 373, 251,
  \dodoi{10.1051/0004-6361:20010632}

\bibitem[{{Edelson} \& {Krolik}(1988)}]{EK88}
{Edelson}, R.~A., \& {Krolik}, J.~H. 1988, \apj, 333, 646,
  \dodoi{10.1086/166773}

\bibitem[{{Egron} {et~al.}(2017){Egron}, {Pellizzoni}, {Giroletti}, {Righini},
  {Stagni}, {Orlati}, {Migoni}, {Melis}, {Concu}, {Barbas}, {Buttaccio},
  {Cassaro}, {De Vicente}, {Gawro{\'n}ski}, {Lindqvist}, {Maccaferri},
  {Stanghellini}, {Wolak}, {Yang}, {Navarrini}, {Loru}, {Pilia}, {Bachetti},
  {Iacolina}, {Buttu}, {Corbel}, {Rodriguez}, {Markoff}, {Wilms},
  {Pottschmidt}, {Cadolle Bel}, {Kalemci}, {Belloni}, {Grinberg}, {Marongiu},
  {Vargiu}, \& {Trois}}]{Egron17}
{Egron}, E., {Pellizzoni}, A., {Giroletti}, M., {et~al.} 2017, \mnras, 471,
  2703, \dodoi{10.1093/mnras/stx1730}

\bibitem[{{Fender} {et~al.}(2004){Fender}, {Belloni}, \& {Gallo}}]{FBG04}
{Fender}, R.~P., {Belloni}, T.~M., \& {Gallo}, E. 2004, \mnras, 355, 1105,
  \dodoi{10.1111/j.1365-2966.2004.08384.x}

\bibitem[{{Fender} {et~al.}(2010){Fender}, {Gallo}, \& {Russell}}]{Fender10}
{Fender}, R.~P., {Gallo}, E., \& {Russell}, D. 2010, \mnras, 406, 1425,
  \dodoi{10.1111/j.1365-2966.2010.16754.x}

\bibitem[{{Fender} {et~al.}(2009){Fender}, {Homan}, \& {Belloni}}]{Fender09}
{Fender}, R.~P., {Homan}, J., \& {Belloni}, T.~M. 2009, \mnras, 396, 1370,
  \dodoi{10.1111/j.1365-2966.2009.14841.x}

\bibitem[{{Fender} {et~al.}(2000){Fender}, {Pooley}, {Durouchoux}, {Tilanus},
  \& {Brocksopp}}]{Fender00}
{Fender}, R.~P., {Pooley}, G.~G., {Durouchoux}, P., {Tilanus}, R.~P.~J., \&
  {Brocksopp}, C. 2000, \mnras, 312, 853,
  \dodoi{10.1046/j.1365-8711.2000.03219.x}

\bibitem[{{Gierli{\'n}ski} \& {Done}(2004)}]{GD04}
{Gierli{\'n}ski}, M., \& {Done}, C. 2004, \mnras, 347, 885,
  \dodoi{10.1111/j.1365-2966.2004.07266.x}

\bibitem[{{Gierli{\'n}ski} \& {Zdziarski}(2005)}]{GZ05}
{Gierli{\'n}ski}, M., \& {Zdziarski}, A.~A. 2005, \mnras, 363, 1349,
  \dodoi{10.1111/j.1365-2966.2005.09527.x}

\bibitem[{{Gierli{\'n}ski} {et~al.}(2010){Gierli{\'n}ski}, {Zdziarski}, \&
  {Done}}]{Gierlinski10}
{Gierli{\'n}ski}, M., {Zdziarski}, A.~A., \& {Done}, C. 2010, Italian Physical
  Soc. Conf. Proc., 103, 229, arXiv:1011.5840

\bibitem[{{Gierlinski} {et~al.}(1997){Gierlinski}, {Zdziarski}, {Done},
  {Johnson}, {Ebisawa}, {Ueda}, {Haardt}, \& {Phlips}}]{G97}
{Gierlinski}, M., {Zdziarski}, A.~A., {Done}, C., {et~al.} 1997, \mnras, 288,
  958

\bibitem[{{Gierli{\'n}ski} {et~al.}(1999){Gierli{\'n}ski}, {Zdziarski},
  {Poutanen}, {Coppi}, {Ebisawa}, \& {Johnson}}]{Gierlinski99}
{Gierli{\'n}ski}, M., {Zdziarski}, A.~A., {Poutanen}, J., {et~al.} 1999,
  \mnras, 309, 496, \dodoi{10.1046/j.1365-8711.1999.02875.x}

\bibitem[{{Grinberg} {et~al.}(2013){Grinberg}, {Hell}, {Pottschmidt},
  {B{\"o}ck}, {Nowak}, {Rodriguez}, {Bodaghee}, {Cadolle Bel}, {Case}, {Hanke},
  {K{\"u}hnel}, {Markoff}, {Pooley}, {Rothschild}, {Tomsick}, {Wilson-Hodge},
  \& {Wilms}}]{Grinberg13}
{Grinberg}, V., {Hell}, N., {Pottschmidt}, K., {et~al.} 2013, \aap, 554, A88,
  \dodoi{10.1051/0004-6361/201321128}

\bibitem[{{Hickish} {et~al.}(2018){Hickish}, {Razavi-Ghods}, {Perrott},
  {Titterington}, {Carey}, {Scott}, {Grainge}, {Scaife}, {Alexander},
  {Saunders}, {Crofts}, {Javid}, {Rumsey}, {Jin}, {Ely}, {Shaw}, {Northrop},
  {Pooley}, {D'Alessand ro}, {Doherty}, \& {Willatt}}]{Hickish18}
{Hickish}, J., {Razavi-Ghods}, N., {Perrott}, Y.~C., {et~al.} 2018, \mnras,
  475, 5677, \dodoi{10.1093/mnras/sty074}

\bibitem[{{Islam} \& {Zdziarski}(2018)}]{Islam18}
{Islam}, N., \& {Zdziarski}, A.~A. 2018, \mnras, 481, 4513,
  \dodoi{10.1093/mnras/sty2597}

\bibitem[{{Kalemci} {et~al.}(2013){Kalemci}, {Din{\c{c}}er}, {Tomsick},
  {Buxton}, {Bailyn}, \& {Chun}}]{Kalemci13}
{Kalemci}, E., {Din{\c{c}}er}, T., {Tomsick}, J.~A., {et~al.} 2013, \apj, 779,
  95, \dodoi{10.1088/0004-637X/779/2/95}

\bibitem[{{Koljonen} \& {Russell}(2019)}]{Koljonen19}
{Koljonen}, K. I.~I., \& {Russell}, D.~M. 2019, \apj, 871, 26,
  \dodoi{10.3847/1538-4357/aaf38f}

\bibitem[{{Kubota} \& {Done}(2016)}]{Kubota_Done16}
{Kubota}, A., \& {Done}, C. 2016, \mnras, 458, 4238,
  \dodoi{10.1093/mnras/stw585}

\bibitem[{{Levine} {et~al.}(1996){Levine}, {Bradt}, {Cui}, {Jernigan},
  {Morgan}, {Remillard}, {Shirey}, \& {Smith}}]{ASM}
{Levine}, A.~M., {Bradt}, H., {Cui}, W., {et~al.} 1996, \apjl, 469, L33,
  \dodoi{10.1086/310260}

\bibitem[{{Liska} {et~al.}(2019){Liska}, {Tchekhovskoy}, {Ingram}, \& {van der
  Klis}}]{Liska19}
{Liska}, M., {Tchekhovskoy}, A., {Ingram}, A., \& {van der Klis}, M. 2019,
  \mnras, 487, 550, \dodoi{10.1093/mnras/stz834}

\bibitem[{{Liska} {et~al.}(2018){Liska}, {Tchekhovskoy}, \&
  {Quataert}}]{Liska18}
{Liska}, M.~T.~P., {Tchekhovskoy}, A., \& {Quataert}, E. 2018, arXiv e-prints.
\newblock \doarXiv{1809.04608}

\bibitem[{{Lubow} {et~al.}(1994){Lubow}, {Papaloizou}, \& {Pringle}}]{Lubow94}
{Lubow}, S.~H., {Papaloizou}, J.~C.~B., \& {Pringle}, J.~E. 1994, \mnras, 267,
  235, \dodoi{10.1093/mnras/267.2.235}

\bibitem[{{Matsuoka} {et~al.}(2009){Matsuoka}, {Kawasaki}, {Ueno}, {Tomida},
  {Kohama}, {Suzuki}, {Adachi}, {Ishikawa}, {Mihara}, {Sugizaki}, {Isobe},
  {Nakagawa}, {Tsunemi}, {Miyata}, {Kawai}, {Kataoka}, {Morii}, {Yoshida},
  {Negoro}, {Nakajima}, {Ueda}, {Chujo}, {Yamaoka}, {Yamazaki}, {Nakahira},
  {You}, {Ishiwata}, {Miyoshi}, {Eguchi}, {Hiroi}, {Katayama}, \&
  {Ebisawa}}]{Matsuoka09}
{Matsuoka}, M., {Kawasaki}, K., {Ueno}, S., {et~al.} 2009, \pasj, 61, 999,
  \dodoi{10.1093/pasj/61.5.999}

\bibitem[{{McKinney} {et~al.}(2012){McKinney}, {Tchekhovskoy}, \& {Bland
  ford}}]{McKinney12}
{McKinney}, J.~C., {Tchekhovskoy}, A., \& {Bland ford}, R.~D. 2012, \mnras,
  423, 3083, \dodoi{10.1111/j.1365-2966.2012.21074.x}

\bibitem[{{Miller-Jones} {et~al.}(2012){Miller-Jones}, {Sivakoff},
  {Altamirano}, {Coriat}, {Corbel}, {Dhawan}, {Krimm}, {Remillard}, {Rupen},
  {Russell}, {Fender}, {Heinz}, {K{\"o}rding}, {Maitra}, {Markoff}, {Migliari},
  {Sarazin}, \& {Tudose}}]{Miller-Jones12}
{Miller-Jones}, J.~C.~A., {Sivakoff}, G.~R., {Altamirano}, D., {et~al.} 2012,
  \mnras, 421, 468, \dodoi{10.1111/j.1365-2966.2011.20326.x}

\bibitem[{{Narayan} {et~al.}(2003){Narayan}, {Igumenshchev}, \&
  {Abramowicz}}]{Narayan03}
{Narayan}, R., {Igumenshchev}, I.~V., \& {Abramowicz}, M.~A. 2003, \pasj, 55,
  L69, \dodoi{10.1093/pasj/55.6.L69}

\bibitem[{{Narayan} \& {McClintock}(2012)}]{Narayan12}
{Narayan}, R., \& {McClintock}, J.~E. 2012, \mnras, 419, L69,
  \dodoi{10.1111/j.1745-3933.2011.01181.x}

\bibitem[{{Pooley} \& {Fender}(1997)}]{PF97}
{Pooley}, G.~G., \& {Fender}, R.~P. 1997, \mnras, 292, 925,
  \dodoi{10.1093/mnras/292.4.925}

\bibitem[{{Poutanen} {et~al.}(2008){Poutanen}, {Zdziarski}, \&
  {Ibragimov}}]{Poutanen08}
{Poutanen}, J., {Zdziarski}, A.~A., \& {Ibragimov}, A. 2008, \mnras, 389, 1427,
  \dodoi{10.1111/j.1365-2966.2008.13666.x}

\bibitem[{{Rodriguez} {et~al.}(1995){Rodriguez}, {Gerard}, {Mirabel}, {Gomez},
  \& {Velazquez}}]{Rodriguez95}
{Rodriguez}, L.~F., {Gerard}, E., {Mirabel}, I.~F., {Gomez}, Y., \&
  {Velazquez}, A. 1995, \apjs, 101, 173, \dodoi{10.1086/192236}

\bibitem[{{Rushton} {et~al.}(2010){Rushton}, {Miller-Jones}, {Paragi},
  {Maccarone}, {Pooley}, {Tudose}, {Fender}, {Spencer}, {Dhawan}, \&
  {Garrett}}]{Rushton10}
{Rushton}, A., {Miller-Jones}, J.~C.~A., {Paragi}, Z., {et~al.} 2010, in 10th
  European VLBI Network Symposium and EVN Users Meeting: VLBI and the New
  Generation of Radio Arrays, Vol.~10, 61

\bibitem[{{Rushton} {et~al.}(2012){Rushton}, {Miller-Jones}, {Campana},
  {Evangelista}, {Paragi}, {Maccarone}, {Pooley}, {Tudose}, {Fender},
  {Spencer}, \& {Dhawan}}]{Rushton12}
{Rushton}, A., {Miller-Jones}, J.~C.~A., {Campana}, R., {et~al.} 2012, \mnras,
  419, 3194, \dodoi{10.1111/j.1365-2966.2011.19959.x}

\bibitem[{{Shapiro} \& {Lightman}(1976)}]{SL76}
{Shapiro}, S.~L., \& {Lightman}, A.~P. 1976, \apj, 204, 555,
  \dodoi{10.1086/154203}

\bibitem[{{Shapopi} \& {Zdziarski}(2020)}]{Shapopi20}
{Shapopi}, J. N.~S., \& {Zdziarski}, A.~A. 2020, arXiv e-prints,
  arXiv:2004.11786.
\newblock \doarXiv{2004.11786}

\bibitem[{{Stirling} {et~al.}(2001){Stirling}, {Spencer}, {de la Force},
  {Garrett}, {Fender}, \& {Ogley}}]{Stirling01}
{Stirling}, A.~M., {Spencer}, R.~E., {de la Force}, C.~J., {et~al.} 2001,
  \mnras, 327, 1273, \dodoi{10.1046/j.1365-8711.2001.04821.x}

\bibitem[{{Tchekhovskoy} {et~al.}(2011){Tchekhovskoy}, {Narayan}, \&
  {McKinney}}]{Tchekhovskoy11}
{Tchekhovskoy}, A., {Narayan}, R., \& {McKinney}, J.~C. 2011, \mnras, 418, L79,
  \dodoi{10.1111/j.1745-3933.2011.01147.x}

\bibitem[{{Wen} {et~al.}(1999){Wen}, {Cui}, {Levine}, \& {Bradt}}]{Wen99}
{Wen}, L., {Cui}, W., {Levine}, A.~M., \& {Bradt}, H.~V. 1999, \apj, 525, 968,
  \dodoi{10.1086/307917}

\bibitem[{{Zamaninasab} {et~al.}(2014){Zamaninasab}, {Clausen-Brown},
  {Savolainen}, \& {Tchekhovskoy}}]{Zamaninasab14}
{Zamaninasab}, M., {Clausen-Brown}, E., {Savolainen}, T., \& {Tchekhovskoy}, A.
  2014, \nat, 510, 126, \dodoi{10.1038/nature13399}

\bibitem[{{Zdziarski}(2012)}]{Zdziarski12}
{Zdziarski}, A.~A. 2012, \mnras, 422, 1750,
  \dodoi{10.1111/j.1365-2966.2012.20754.x}

\bibitem[{{Zdziarski} \& {Gierli{\'n}ski}(2004)}]{ZG04}
{Zdziarski}, A.~A., \& {Gierli{\'n}ski}, M. 2004, Progr. Theor. Phys. Suppl.,
  155, 99, \dodoi{10.1143/PTPS.155.99}

\bibitem[{{Zdziarski} {et~al.}(2017){Zdziarski}, {Malyshev}, {Chernyakova}, \&
  {Pooley}}]{ZMC17}
{Zdziarski}, A.~A., {Malyshev}, D., {Chernyakova}, M., \& {Pooley}, G.~G. 2017,
  \mnras, 471, 3657.
\newblock \doarXiv{1607.05059}

\bibitem[{{Zdziarski} {et~al.}(2011{\natexlab{a}}){Zdziarski}, {Pooley}, \&
  {Skinner}}]{ZPS11}
{Zdziarski}, A.~A., {Pooley}, G.~G., \& {Skinner}, G.~K. 2011{\natexlab{a}},
  \mnras, 412, 1985, \dodoi{10.1111/j.1365-2966.2010.18034.x}

\bibitem[{{Zdziarski} {et~al.}(2011{\natexlab{b}}){Zdziarski}, {Skinner},
  {Pooley}, \& {Lubi{\'n}ski}}]{ZSPL11}
{Zdziarski}, A.~A., {Skinner}, G.~K., {Pooley}, G.~G., \& {Lubi{\'n}ski}, P.
  2011{\natexlab{b}}, \mnras, 416, 1324,
  \dodoi{10.1111/j.1365-2966.2011.19127.x}

\bibitem[{{Zdziarski} {et~al.}(2018){Zdziarski}, {Malyshev}, {Dubus}, {Pooley},
  {Johnson}, {Frankowski}, {De Marco}, {Chernyakova}, \& {Rao}}]{Z18}
{Zdziarski}, A.~A., {Malyshev}, D., {Dubus}, G., {et~al.} 2018, \mnras, 479,
  4399, \dodoi{10.1093/mnras/sty1618}

\bibitem[{{Zi{\'o}{\l}kowski}(2014)}]{Ziolkowski14}
{Zi{\'o}{\l}kowski}, J. 2014, \mnras, 440, L61, \dodoi{10.1093/mnrasl/slu002}

\bibitem[{{Zwart} {et~al.}(2008){Zwart}, {Barker}, {Biddulph}, {Bly}, {Boysen},
  {Brown}, {Clementson}, {Crofts}, {Culverhouse}, {Czeres}, {Dace}, {Davies},
  {D'Alessandro}, {Doherty}, {Duggan}, {Ely}, {Felvus}, {Feroz}, {Flynn},
  {Franzen}, {Geisb{\"u}sch}, {G{\'e}nova-Santos}, {Grainge}, {Grainger},
  {Hammett}, {Hills}, {Hobson}, {Holler}, {Hurley-Walker}, {Jilley}, {Jones},
  {Kaneko}, {Kneissl}, {Lancaster}, {Lasenby}, {Marshall}, {Newton}, {Norris},
  {Northrop}, {Odell}, {Petencin}, {Pober}, {Pooley}, {Pospieszalski}, {Quy},
  {Rodr{\'\i}guez-Gonz{\'a}lvez}, {Saunders}, {Scaife}, {Schofield}, {Scott},
  {Shaw}, {Shimwell}, {Smith}, {Taylor}, {Titterington}, {Veli{\'c}},
  {Waldram}, {West}, {Wood}, {Yassin}, \& {AMI Consortium}}]{Zwart08}
{Zwart}, J.~T.~L., {Barker}, R.~W., {Biddulph}, P., {et~al.} 2008, \mnras, 391,
  1545, \dodoi{10.1111/j.1365-2966.2008.13953.x}

\end{thebibliography}

\label{lastpage}
\end{document}